\documentstyle[prl,aps,epsfig]{revtex}
\begin{document}
\draft
\twocolumn[\hsize\textwidth\columnwidth\hsize\csname
@twocolumnfalse\endcsname
\title{ Effect of Spinless Impurities on Reduction of $T_c$ in High $T_c$
 Superconductors}
\author{ In-Ho Lee$^1$ and Sang Boo Nam$^2$\cite{sbnam}  \\}
\address{$^1$School of Physics, Korea Institute for Advanced Study,
Cheongryangri-dong, Dongdaemun-gu, Seoul 130-012, Korea \\
$^2$Superconductivity Group, Korea Research Institute of Standards and Science,\\
     Doryong-dong~ 1, Yusung-ku, Taejeon 305-600, Korea}
\maketitle
\begin{abstract}
The notion of a finite pairing interaction energy range suggested by Nam, results in 
some states at the Fermi level not participating in pairings when there are scattering
centers such as impurities. The fact that not all states at the Fermi level participate
in pairing is shown to suppress $T_c$ in an isotropic superconductor and destroy
superconductivity. We have presented quantitative calculations of 
$T_c$ reduced via spinless impurities, in good agreements with data of Zn-doped YBCO 
and LSCO, respectively. It is not necessary to have 
the anisotropic order parameter,
to account for the destruction of superconductivity via non-magnetic impurities.
\end{abstract}
]

One of intriguing experiments in high $T_c$ superconductors (HTS) is that the non-magnetic
impurity Zn reduces $T_c$ and  destroy superconductivity in Zn-doped YBCO\cite{1,2,3}
and Zn-doped LSCO\cite{4}, respectively.

The spinless impurity scattering, according to Anderson\cite{5}, would not change 
thermodynamical properties of superconductor, such as $T_c$, providing the order parameter
being a uniform function, that is, a constant. He argued that the pairing of the state
and its time reversal state in the scattering quantum space, would yield the same order
parameter as that in the unscattered case, since the states in the former space can be
mapped via the unitary transformation from those in the latter. In practice, one of crucial
parameters to determine $T_c$ is the BCS coupling parameter
\begin{equation}
g=\langle N(0) V_{BCS}  \rangle
\end{equation}
with the density of states at the Fermi level $N(0)$ and the BCS pairing interaction $V_{BCS}$
averaged. We expect, thus, that the anisotropic pairing interaction would affect $T_c$.
In fact, Markowitz and Kadanoff\cite{6} have examined the anisotropic effect on $T_c$ and 
obtained good agreements between calculations and data for low concentrations of non-magnetic
impurities in low $T_c$ superconductors (LTS).

On the other hand, Abrikosov and Gor'kov (AG)\cite{7} have shown that the spin-flip 
scatterings suppress $T_c$ and destroy superconductivity. Also, when the order parameter 
has nodes, say $p$- or $d$-wave pairing states, 
Hirschfeld,  W\"olfle, and  Einzel\cite{8}
argued that the resonance impurity scattering causes $T_c$ be reduced in the way of AG
formula. In essence, the anisotropic effects and pair breaking
scatterings, such as, spin-flip scattering, are known to be, up to now,
the causes for destruction of superconductivity. We will show shortly a new chapter of
destruction of superconductivity via non-magnetic impurity scatterings in an isotropic
superconductor.

The notion of a finite pairing interaction energy range $T_d$\cite{9} results in the
incomplete condensation in which not all states participate in pairings.
The states not participating in pairings yield low energy states responsible for the linear
$T$ dependence of superelectron density at low $T$ in an isotropic superconductor\cite{10}.
The quantitative calculations of magnetic penetration depth length in all $T$ ranges based on a 
finite $T_d$\cite{11} are in good agreements with data of YBCO\cite{12,13}, BSCCO\cite{14},
HBCCO\cite{15}, LSCO\cite{16}, and Sr214\cite{17}, respectively.
Moreover, the incomplete condensation results in the picture of
multiconnected superconductors\cite{18} which can account for the $\pi$-phase shift in the
Pb-YBCO SQUID\cite{19} and $\frac{1}{2}$ magnetic fluxoid quantum observed in the YBCO ring
with three grain boundary junctions\cite{20}.

First we recapitulate the pertinent results of a finite $T_d$\cite{9}. To see the phase 
transition, $T_c$ should have a finite value, that is, neither zero nor infinite. 
To have a finite value $T_c$, $T_d$ should be finite, since $T_c$ is scaled with $T_d$
within the pairing theory. In other words, we may write the order parameter 
$\Delta(k,\omega)$ as\cite{9}
\begin{eqnarray}
\Delta(k, \omega)=\left\{
\begin{array}{ll}
\Delta  &{\rm~~~for}~~ |\epsilon_k| <T_d \\
 0      &{\rm~~~for}~~ |\epsilon_k| > T_d
\end{array}
\right\}
\label{DELTA}
\end{eqnarray}
for all dynamical energies $\omega$. Here $\epsilon_k$ is the usual normal state
excitation energy of wave number $k$, measured with respect to the Fermi energy. 
Hereafter units of $\hbar=c=k_B=1$ are used. And later $\Delta(T)$ for $\Delta$
may be used as well. The $\Delta$ is a solution of BCS like equation
\begin{equation}
\frac{1}{g}=\int_0^{T_d} \frac{d\epsilon}{E}
\tanh\left ( \frac{ E  }{2T} \right ).
\label{GAPEQ}
\end{equation}
with $E=(\epsilon^2+\Delta^2)^{1/2}$.
For $\Delta(T_{c0})=0$, we get\cite{9}
\begin{equation}
1/g=(2/\pi) \sum_j (2/j) \tan^{-1}(y/j),
\label{NAMTC}
\end{equation}
where $y=T_d/(\pi T_{c0})$ with the transition temperature $T_{c0}$ for a pure system,
and sum is over the positive odd integers $j$. The factor of arctangent function makes
the sum converge. For large $y$, Eq. (\ref{NAMTC}) yields the BCS result $T_{c0}$(BCS).
The quantitative calculations of $T_{c0}$ and $\Delta$ vs $g$ and $T$ 
are given in Ref. \cite{11}.
The $T_{c0}$ from Eq. (\ref{NAMTC}) is always greater than $T_{c0}$(BCS)
and does not have any upper limit.
The fact is that for large $g > 2.32$, $T_{c0}$ increases with increasing $g$ as 
$T_{c0}\approx gT_d/2$.  One interesting value of $g=0.657$ yields $T_{c0}\approx100$ K with
$T_{d}=400$ K which is of the order of Debye temperature in HTS. This value of $g$ may be
realized in YBCO by considering the electron-phonon interaction of the order of 
$\lambda_p=1.3 \sim 2.3$\cite{21}.

The key idea of this letter is a novel one. When there are scattering centers such as
impurities in a Fermi system, the spectral weight distribution function at the Fermi level,
that is, the imaginary part of the Green's function at $\omega=0$, may be considered as a
Lorentzian form.  The sum of spectral weights outside $T_d < |\epsilon_k|$, that is,
the integral of the imaginary part of the Green's function with $\epsilon_k$ in the
ranges of $|\epsilon_k| > T_d$, is given as  
\begin{equation}
N(0)R=N(0)(2/\pi)\tan^{-1}(\Gamma/T_d),
\end{equation}
where $\Gamma$ is the imaginary part of self-energy due to the impurity scatterings, or
the half impurity scattering rate $1/(2\tau)$. The states of $N(0)R$ do not participate
in pairings, and are predicted\cite{23} to be responsible for the linear $T$ dependence 
of the specific heat at low $T$ in Zn-doped YBCO\cite{24,25}
and also for the $T^2$ term in the magnetic penetration depth length 
at low $T$ in the imperfect HTS such as Sr214\cite{17}.
The effective density of states at the Fermi level participating in pairings,
becomes $N(0)(1-R)$ less than the usual one $N(0)$, and results in the reduced $T_c$.
For simplicity, we assumed here the effective BCS coupling parameter $g$ be not changed
via impurity scatterings. Also the Born approximation is used for calculations of impurity
scatterings.

To incorporate the above key idea into the calculation of $T_c$, 
we use the Green's function formalism\cite{22}.
In the thermal Green's function scheme,
i.e., replacing $\omega$ by $i\omega_n=i(2n+1)\pi T$ with integer $n$,
we may, taking into account the order parameter of Eq. (\ref{DELTA}) consistently, 
rewrite Eqs. (3.5a, b, c) of Ref. \cite{22} as
\begin{eqnarray}
Z_n \omega_n &=&\omega_n  + (\Gamma+\Gamma_s) [\Phi_1(\omega_n)+\Phi_2(\omega_n)],
 \label{GDELTA1} \\
\nonumber \\
Z_n \Delta_n &=&\Delta+ (\Gamma-\Gamma_s)\Phi_3(\omega_n), 
\label{GDELTA2} \\
\nonumber \\
\Delta       &=& gT\sum_n \Phi_3(\omega_n),   \label{GDELTA3}
\end{eqnarray}
\begin{eqnarray}     
\Phi_1(\omega_n)  &=& (\omega_n/E_n) (2/\pi) \tan^{-1}[T_d/(Z_n E_n)],  \nonumber \\
\nonumber \\
\Phi_2(\omega_n)  &=& (2/\pi)\tan^{-1}(Z_n \omega_n/T_d),  \nonumber \\
\nonumber \\
\Phi_3(\omega_n)  &=& (\Delta_n/\omega_n)\Phi_1(\omega_n),  \nonumber \\
\nonumber \\
E_n       &=& (\omega_n^2+\Delta^2)^{1/2}.  \nonumber 
\end{eqnarray}
Here $\Gamma$ and  $\Gamma_s$ are the imaginary parts of self-energy via non-magnetic 
and magnetic impurity scatterings, respectively. When $\Gamma=0=\Gamma_s$, $Z_n=1$,
and $\Delta_n=\Delta$, Eq. (\ref{GDELTA3}) becomes Eq. (\ref{GAPEQ}) 
of the BCS like equation, as it should.

For $T \rightarrow  T_c$, $\Delta \rightarrow 0$, from Eqs. (\ref{GDELTA1}), 
(\ref{GDELTA2}), and (\ref{GDELTA3}), we, after simple algebras, get the equation for 
$T_c$ as\cite{23}
\begin{equation}
\frac{1}{g} = \sum_n \frac{2\pi T_c r_n}{\omega_n+D_n},
\label{NTC}
\end{equation}
\begin{eqnarray}     
D_n &=& (1-r_n)\Gamma+(1+r_n)\Gamma_s,  \nonumber  \\
\nonumber \\
r_n &=& (2/\pi)\tan^{-1}[T_d/(\omega_n+\Gamma+\Gamma_s)], \nonumber   \\
\nonumber \\
\omega_n &=& (2n+1)\pi T_c. \nonumber
\end{eqnarray}
The sum is over integers $n \ge 0$. For $\Gamma=0=\Gamma_s$, Eq. (\ref{NTC})
becomes Eq. (\ref{NAMTC}) for a pure superconductor. In the infinite $T_d$ limit,
equivalently small $g$ limit, that is, LTS case, combining 
Eq. (\ref{NAMTC}) and  Eq. (\ref{NTC})  yields the AG result, with $r_n=1$ and
$D_n=2\Gamma_s$, as \cite{7,22}
\begin{equation}
\ln(T_c/T_{c0})=\sum_n \left (\frac{2}{2n+1+d}-\frac{2}{2n+1} \right ),
\end{equation}
where the depairing parameter $d=2\Gamma_s/(\pi T_c)$.
Furthermore, $T_c$ becomes independent of $\Gamma$, and the Anderson
theorem\cite{5} follows with $\Gamma_s=0$. 
In general cases, combining Eq. (\ref{NAMTC}) and 
Eq. (\ref{NTC}), we calculate easily $T_c/T_{c0}$ and 
$T_c/T_{d}$ vs $\Gamma$,  $\Gamma_s$, and $g$, respectively.
The calculations for $g=0.3$, 0.45, 0.657, 1, and 2, have been carried out, 
and some results are presented here.

In Fig. 1 is shown $T_c/T_d$ vs $2\Gamma/T_d$ and $2\Gamma_s/T_d$ for 
$g=0.657$ for example.
The general shape of $T_c/T_d$ appears to be similar 
in any direction of $\Gamma$ and $\Gamma_s$. 
\begin{minipage}[H]{0.78\linewidth} \vspace{1.0cm}
\centering\epsfig{file=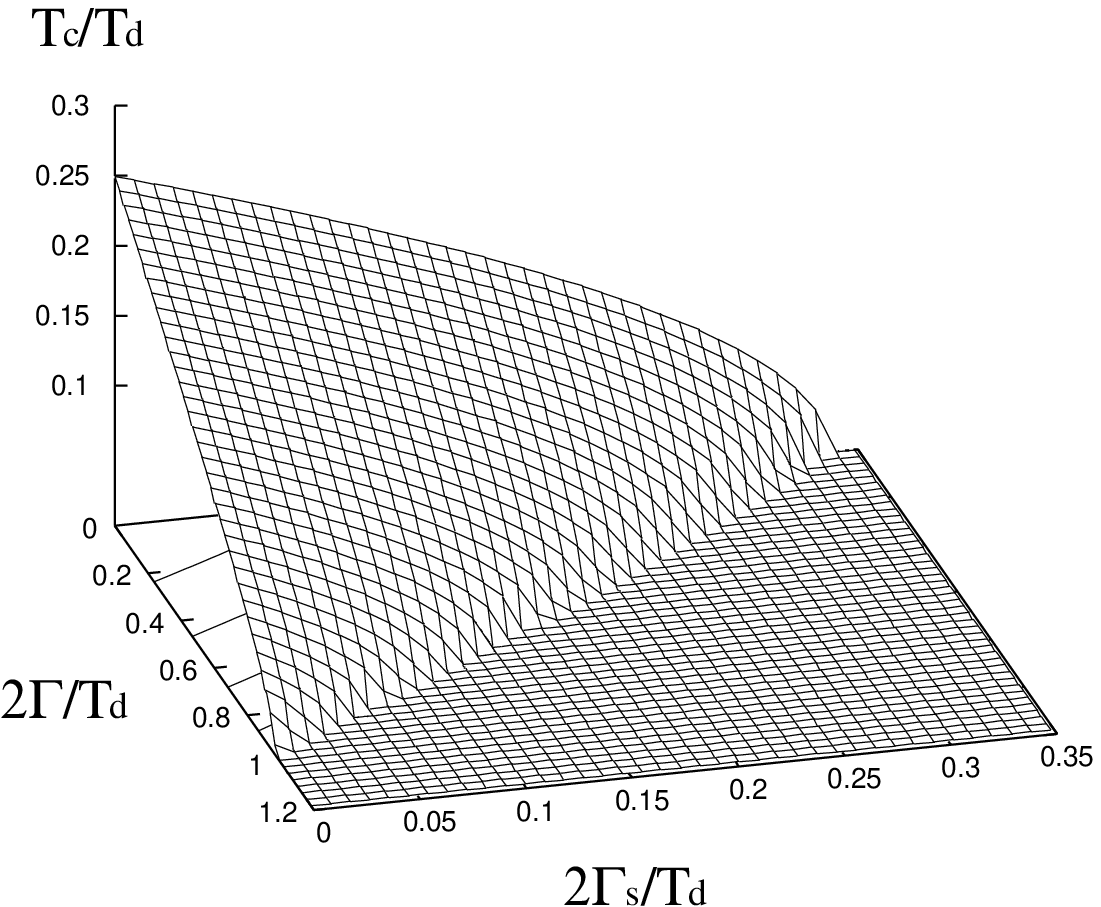, width=1.22\linewidth,angle=0}
\end{minipage} 
\begin{figure}
\caption{ $T_c/T_d$ vs $2\Gamma/T_d$ and $2\Gamma_s/T_d$ for $g=0.657$.}
\end{figure}
\noindent
To see  somewhat in detail, in Fig. 2 are shown $T_c/T_d$ vs $2\Gamma/T_d$ 
for $\Gamma_s=0$ and $T_c/T_d$ vs $2\Gamma_s/T_d$ for $\Gamma=0$,
for $g=0.3$, 0.45, 0.657, 1, and 2, respectively.
\begin{minipage}[H]{0.73\linewidth} \vspace{1.0cm}
\centering\epsfig{file=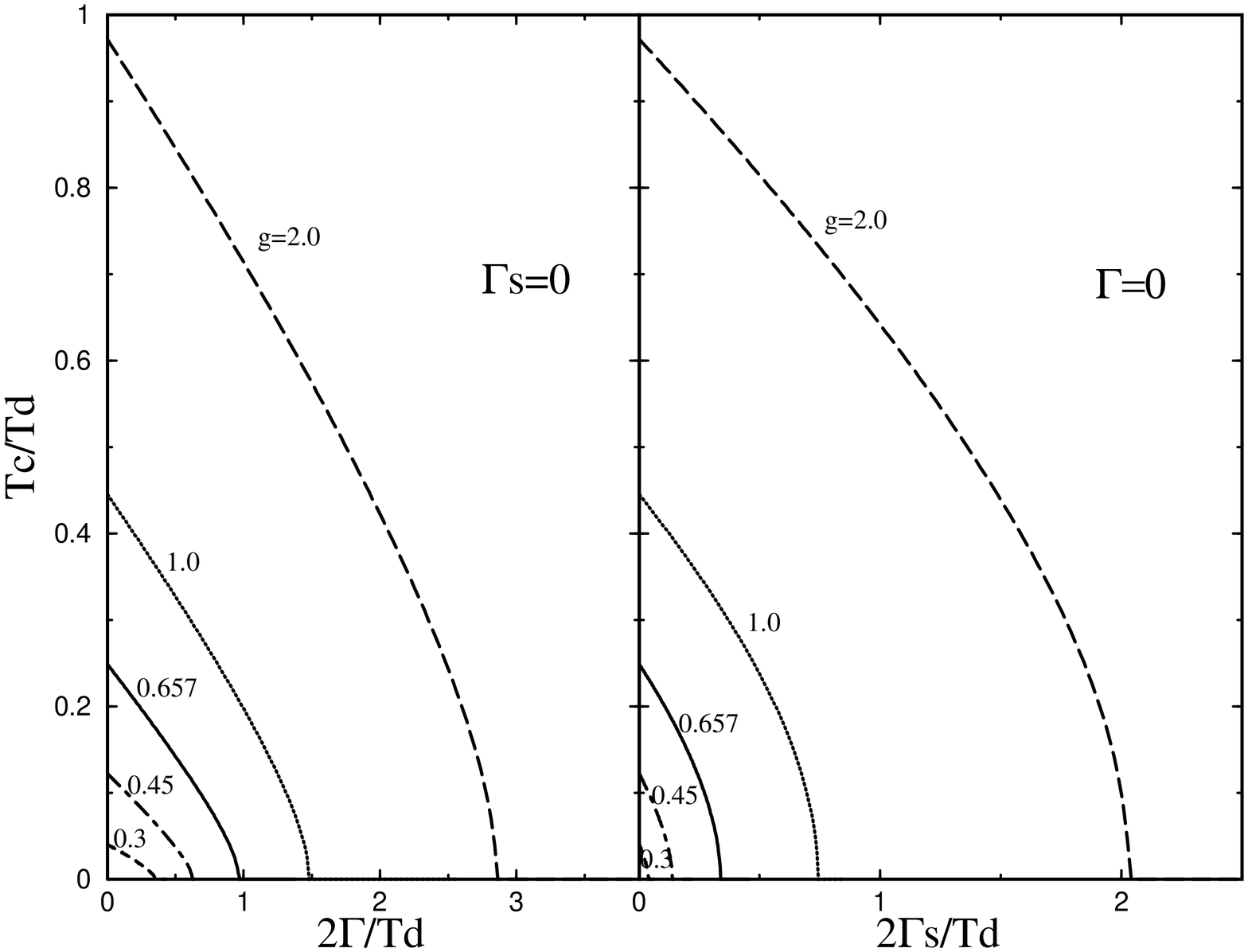, width=1.0\linewidth,angle=270}
\end{minipage} 
\begin{figure}
\caption{ The left panel is $T_c/T_d$ vs $2\Gamma/T_d$ for $\Gamma_s=0$, 
while the right panel is $T_c/T_d$ vs $2\Gamma_s/T_d$ for $\Gamma=0$.}
\end{figure}
\noindent
The reductions of $T_c$ via $\Gamma$ are slow for small $g$, 
and the steepness increases with increasing $g$. On the other hand, the reductions
of $T_c$ via $\Gamma_s$ are almost the same shape for all $g$ values.

To see some more differences between non-magnetic and magnetic impurity scatterings, we
have changed the scales of $2\Gamma/T_d$ and $2\Gamma_s/T_d$ by
                           $2\Gamma/\Delta_0$ and $2\Gamma_s/\Delta_0$, with 
$\Delta_0=\Delta(0)$ the order parameter in a pure superconductor, at $T=0$, and a
function of $g$. In Fig. 3 are shown $T_c/T_d$ vs $2\Gamma/\Delta_0$  and
$T_c/T_d$ vs $2\Gamma_s/\Delta_0$.
\begin{minipage}[H]{0.73\linewidth} \vspace{1.0cm}
\centering\epsfig{file=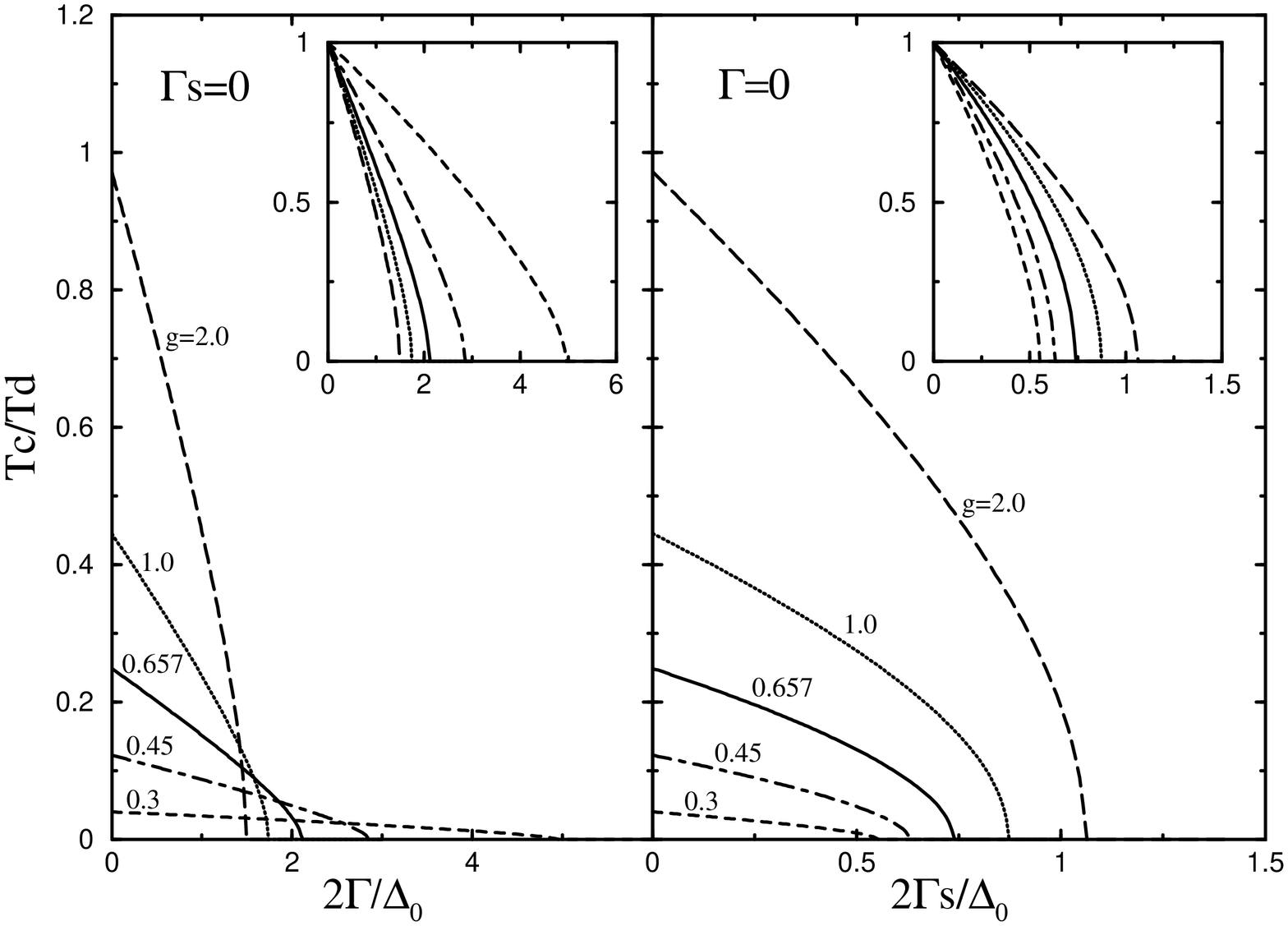, width=0.96\linewidth,angle=270}
\end{minipage} 
\begin{figure}
\caption{ 
The left panel and the right panel are
$T_c/T_d$ vs $2\Gamma/\Delta_0$   with $\Gamma_s=0$ and
$T_c/T_d$ vs $2\Gamma_s/\Delta_0$ with $\Gamma=0$, respectively. 
The normalized $T_c/T_{c0}$ are given in the insets. }
\end{figure}
\noindent
The insets in Fig. 3 contain the normalized
$T_c/T_{c0}$ vs $2\Gamma/\Delta_0$ 
and $T_c/T_{c0}$ vs $2\Gamma_s/\Delta_0$, respectively.  
The normalized $T_c/T_{c0}$ for all cases have almost the same shape. 

In the $T_c=0$ limit, we may rewrite  Eq. (\ref{NTC}) as\cite{23}
\begin{equation}
\frac{1}{g}=\int_0^{\infty} dx \frac{r_c(x)}{x+D_c(x)},
\label{NAMXX}
\end{equation}
\begin{eqnarray}
r_c(x) &=& (2/\pi) \tan^{-1} [(T_d/\Delta_0)/(x+x_1+x_2)],   \nonumber \\
\nonumber \\
D_c(x) &=& [1-r_c(x)] x_1+[1+r_c(x)] x_2,   \nonumber \\
\nonumber \\
x_1    &=& \Gamma_c/\Delta_0,   \nonumber \\
\nonumber \\
x_2    &=& \Gamma_{sc}/\Delta_0.   \nonumber
\end{eqnarray}
The phase diagram for $T_c=0$ is shown in Fig. 4.
\begin{minipage}[H]{0.73\linewidth} \vspace{1.0cm}
\centering\epsfig{file=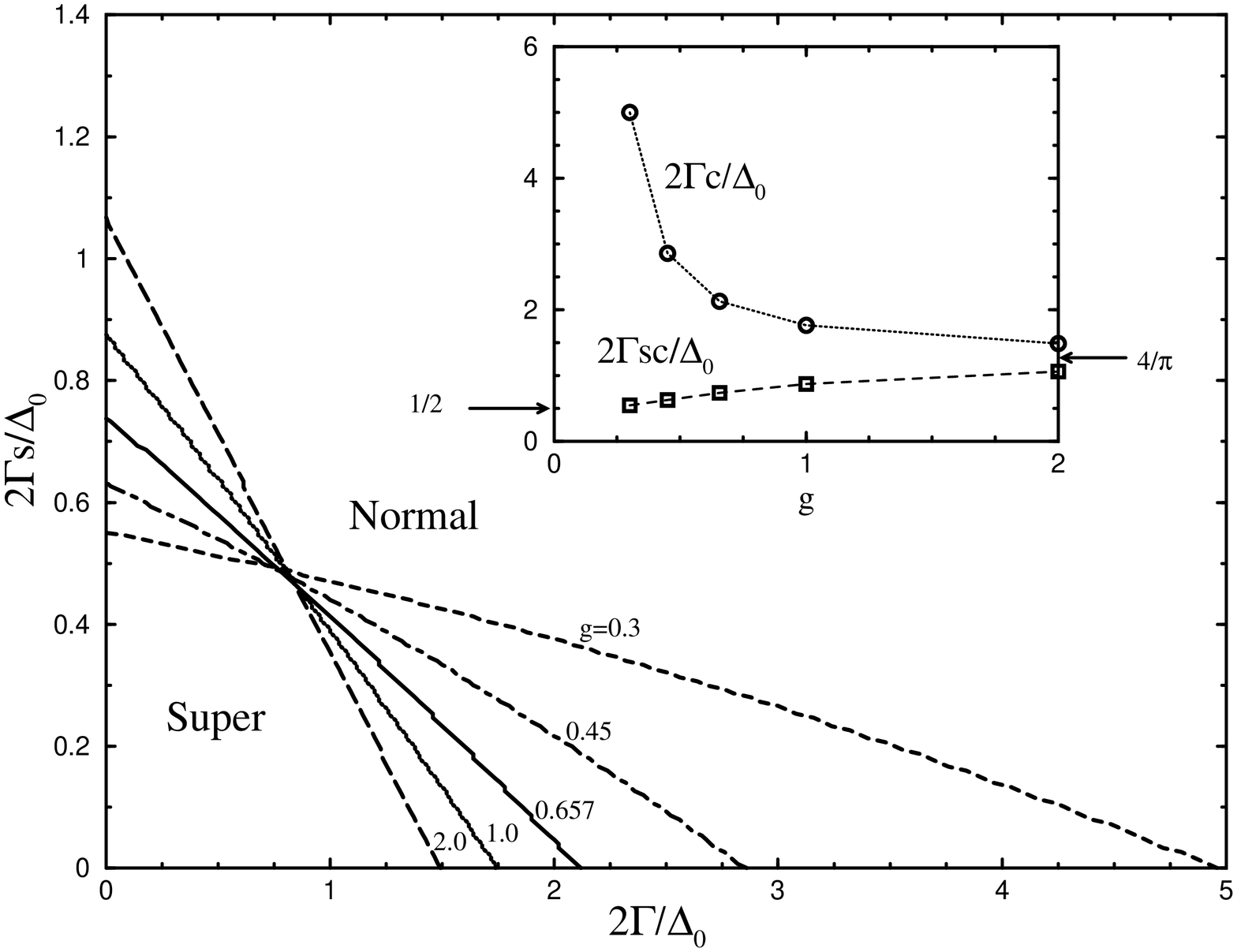, width=0.99\linewidth,angle=270}
\end{minipage} 
\begin{figure}
\caption{ The phase diagram on $\Gamma-\Gamma_s$ plane. The critical values of 
$\Gamma_c$ and $\Gamma_{sc}$ are given in the inset.}
\end{figure}
\noindent
In the inset in Fig. 4, 
the critical vaules of
$2\Gamma_c/\Delta_0$ decrease with increasing $g$, 
contrast to those of $2\Gamma_c/T_d$ in Fig. 2 increase with increasing $g$, 
since $T_c$ increases with increasing $g$.
However, both $2\Gamma_{sc}/\Delta_0$ and $2\Gamma_{sc}/T_d$ increase with increasing $g$.
For small $g$ limit, LTS case, the AG result
$2\Gamma_{sc}/\Delta_0=0.5$ is obtained. In the large $g$ limit, HTS case, we get
from Eq. (\ref{NAMXX}) as \cite{23}
\begin{equation}
2\Gamma_c/\Delta_0+2\Gamma_{sc}/\Delta_0  =4/\pi.
\end{equation}
This can be realized from Eq. (\ref{NTC}) as well by observing $r_n=0$ and
$D_n=\Gamma+\Gamma_s$ for large $g$ or small $T_d$ limit. This implies the effects of 
non-magnetic and magnetic scatterings on $T_c$ are the same in HTS. Data in HTS indeed
indicated such as shown in Fig. 5. To compare calculations with data, we used
$T_c/T_{c0}$ vs $\Gamma/\Gamma_c$ in Fig. 5. In addition to data of Zn-doped HTS, 
those of Ni-doped YBCO and LSCO also 
are included to demonstrate both non-magnetic and magnetic
scatterings affect $T_c$ in the same way for large $g$, that is, HTS case.
\begin{minipage}[H]{0.73\linewidth} \vspace{1.0cm}
\centering\epsfig{file=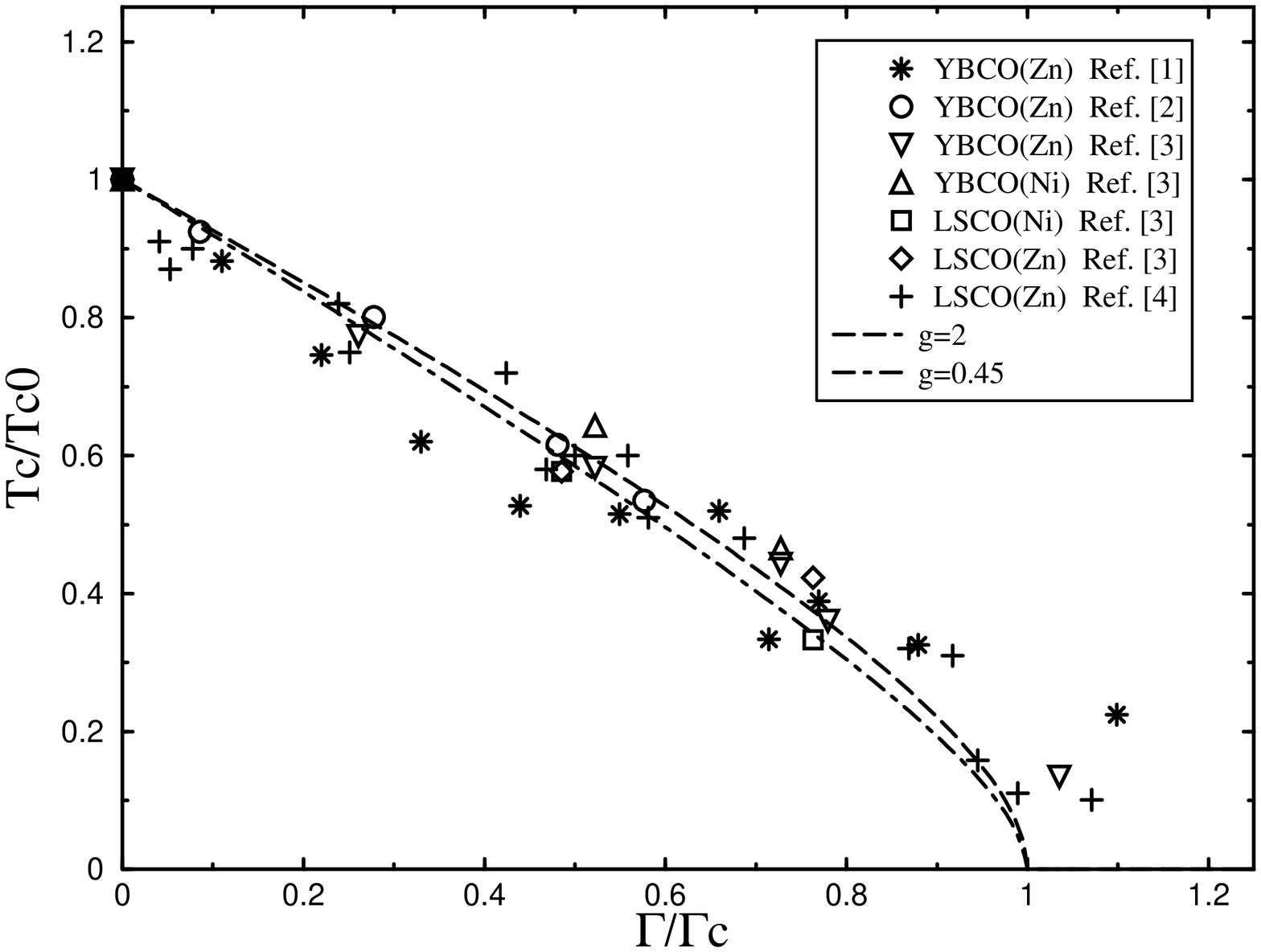, width=1.0\linewidth,angle=270}
\end{minipage} 
\begin{figure}
\caption{Comparison between calculations from Eq. (\ref{NTC}) and
data of Zn-doped and Ni-doped YBCO and LSCO.
Only two theoretical curves ($g=2.0$ and 0.45) are shown. 
Others are bounded by these two curves. }
\end{figure}
\noindent
The data of YBCO\cite{1} are poor but that of YBCO\cite{2,3} 
are in good agreements with our calculations.
The data of LSCO\cite{4} are satisfactory. 
In all, data are understood in the framework of a finite $T_d$.

In summary, even though the model of the order parameter of 
Eq. (\ref{DELTA}) is ideal, and the Born approximation 
of impurity scatterings in an isotropic superconductor is used, 
the calculations of $T_c$ vs non-magnetic and magnetic
impurity scatterings,
are in good agreements with data of YBCO and LSCO, respectively.
The notable results are the normalized $T_c/T_{c0}$ vs $\Gamma/\Gamma_c$ are almost
the same each other for all $g$ values, and 
$T_c/T_{c0}$ vs $\Gamma_s/\Gamma_{sc}$ yields the AG result. For improvement,  one may
use the $T$ matrix formulation for impurity scatterings, and take into account the 
retardation effect on the order parameter. However, we expect the $T_c/T_{c0}$ would not 
be changed much from that discussed here. The spinless impurity scatterings can destroy
superconductivity in an isotropic superconductor as well. The notion of a finite $T_d$
is sound theoretically. We hope it would account for other unsolved problems in HTS.

SBN thanks KRISS members for their warm hospitalities at KRISS, 
Drs. J. C. Park and Y. K. Park for discussions, 
and Drs. J. Lee and T. Kang for their kind helps. This work is supported
in part by KOFST.


\begin{thebibliography}{999}
\bibitem[*]{sbnam} Correspondence address: wonkinam@kriss.re.kr
\bibitem{1} G. Xiao, M. Z. Cieplak, A. Gavrin, F. H. Streitz, A. Bakhshai, and C. L. Chien,
   \prl {\bf 60}, 1446 (1988).
\bibitem{2} T. R. Chien, Z. Z. Wang, and N. P. Ong,  \prl {\bf 67}, 2088 (1991).
\bibitem{3} G. V. M. Williams, J. L. Tallon, and R. Dupree,  \prb {\bf 61}, 4319 (2000); 
  G. V. M. Williams and J. L. Tallon, \prb {\bf 57}, 10984 (1998).  
\bibitem{4} K. Karpiska, M. Z. Cieplak, S. Guha, A. Malinowski, T. Skokiewicz,
 W. Plesiewicz, M.  Berkowski, B. Boyce, T. R. Lemberger, and P. Lindenfeld, 
       \prl {\bf 84}, 155 (2000).
\bibitem{5} P. W. Anderson, J. Phys. Chem. Solids. {\bf 11}, 26 (1959).
\bibitem{6} D. Markowitz and L. P. Kadanoff, Phys. Rev. {\bf 136}, 563 (1963).
\bibitem{7} A. A. Abrikosov and L. P. Gor'kov, JETP {\bf 12}, 1243 (1961).
\bibitem{8} P. J. Hirschfeld,  P. W\"olfle, and D. Einzel, \prb {\bf 37}, 83 (1988).
\bibitem{9} S. B. Nam, Phys. Lett. A {\bf 193}, 111 (1994); ibid(E) A {\bf 197}, 458 (1995).
\bibitem{10} S. B. Nam, Prog. in Supercond. {\bf 2}, 11 (2000).
\bibitem{11} J.-W. Lee, I.-H. Lee, and S. B. Nam, cond-mat/0101011
\bibitem{12} W. N. Hardy, D. A. Bonn, D. C. Morgan, R. Liang, and K. Zhang,
             \prl {\bf 70}, 3999 (1993).
\bibitem{13} S. Kamal, R. Liang, A. Hosseini, D. A. Bonn, and W. N. Hardy,
          \prb {\bf 58}, R8933 (1998).
\bibitem{14} S.-F. Lee, D. C. Morgan, R. J. Ormeno, D. M. Broun, R. A. Doyle, 
 J. R. Waldram, and K. Kadowaki, \prl {\bf 77}, 735 (1996).
\bibitem{15} C. Panagopoulos, J. R. Cooper, G. B. Peacock, I. Gameson, P. P. Edwards, 
 W. Schmidbauer, and J. W. Hodby,  \prb {\bf 53}, R2999 (1996).
\bibitem{16} C. Panagopoulos, B. D. Rainford,  J. R. Cooper, W. Lo,
J. L. Tallon, J. W. Loram, J. Betouras, Y. S. Wang, and C. W. Chu,  
  \prb {\bf 60}, 14617 (1999).
\bibitem{17} I. Bonalde, Brian D. Yanoff, M. B. Salamon, D. J. Van Harlingen,  E. M. E. Chia, 
 Z. Q. Mao, and Y. Maeno, \prl {\bf 85}, 4775 (2000).
\bibitem{18} S. B. Nam, Phys. Lett. A {\bf 198}, 447 (1995).
\bibitem{19} D. A. Wollman, D. J. Van Harlingen, W. C. Lee, D. M. Ginsberg, 
                 and A. J. Leggett, \prl {\bf 71}, 2134 (1993).
\bibitem{20} C. C. Tsuei, J. R. Kirtley, C. C. Chi, Lock See Yu-Jahnes, A. Gupta, T. Shaw, 
             J. Z. Sun, and M. B. Ketchen, \prl {\bf 73}, 593 (1994).
\bibitem{21} W. Weber and  L. F. Mattheiss, \prb {\bf 37}, 599 (1988).
\bibitem{23} S. B. Nam, Bull. Am. Phys. Soc. {\bf 45}, 256 (2000).
\bibitem{24} K. A. Moler, D. L. Sisson, J. S. Urbach, M. R. Beasley, 
    A. Kapitulnik, D. J. Baar, R. Liang, and W. N. Hardy, 
         \prb {\bf 55}, 3954 (1997).
\bibitem{25} D. L. Sisson, S. G. Doettinger, A. Kapitulnik, R. Liang, D. A. Bonn, 
              and W. N. Hardy,  \prb {\bf 61}, 3604 (2000).
\bibitem{22} S. B. Nam, Phys. Rev. {\bf 156}, 470, 487 (1967).
\end{thebibliography}
\end{document}